\documentclass{LT23auth}
\usepackage{graphicx}

\begin{document}

\begin{frontmatter}

\title{Isolated resonances in conductance fluctuations in ballistic billiards}

\author[address1]{Arnd B\"acker},
\author[address2]{Achim Manze \thanksref{thank1}},
\author[address2]{Bodo Huckestein},
\author[address3]{Roland Ketzmerick}

\address[address1]{Abteilung Theoretische Physik, 
  Universit\"at Ulm, Albert-Einstein-Allee 11, D-89081 Ulm, Germany}
\address[address2]{Institut f\"ur Theoretische Physik III, Ruhr-Universit\"at
  Bochum, D-44780 Bochum, Germany}
\address[address3]{Max-Planck-Institut f\"ur Str\"omungsforschung and Institut
  f\"ur Nichtlineare Dynamik der Universit\"at G\"ottingen, Bunsenstra{\ss}e 10, D-37073 G\"ottingen, Germany}

\thanks[thank1]{ E-mail: amz@tp3.ruhr-uni-bochum.de}

\begin{abstract}
  We study the isolated resonances occurring in conductance
  fluctuations of ballistic electron systems with a classically mixed
  phase space. In particular, we calculate the conductance and
  Wigner-Smith time as well as scattering states and eigenstates of
  the open and closed cosine billiard, respectively. We demonstrate
  that the observed isolated resonances and their scattering states
  can be associated with eigenstates of the closed system. They can
  all be categorized as hierarchical or regular, depending on where in
  a phase space representation the corresponding eigenstates are
  concentrated.
\end{abstract}
%
%
\end{frontmatter}

One of the properties in which classical chaos manifests itself
quantum mechanically is the conductance. A well known example is the
occurrence of the universal conductance fluctuations in billiards
whose classical counterparts show completely chaotic
dynamics~\cite{Jal2000}. However, generic systems are neither
completely chaotic nor integrable but have a mixed phase space, where
regions of regular motion coexist with those of chaotic
motion~\cite{LicLie92}.  A semiclassical analysis for the quantum
mechanical analog of such systems showed that the graph of conductance
$G$ vs control parameter should be a fractal \cite{Ket96} (fractal conductance fluctuations).

Surprisingly, for the cosine billiard, a system with a mixed phase
space, a recent numerical study did not show these fractal conductance
fluctuations but instead sharp isolated resonances with a width
distribution covering several orders of magnitude~\cite{HucKetLew99} 
superimposed on universal conductance fluctuations of order unity.
The resonances in the conductance are accompanied by equally sharp but
much stronger resonances in the Wigner-Smith time. We will show that
the origins of these resonances are quantum mechanical states with
phase space portraits that are concentrated on the regular and
hierarchical regions of phase space \cite{KetHufSteWei2000}, in
aggreement with the conjecture of Ref.~\cite{HufWeiKet2001}.

\begin{figure}[btp]
  \begin{center}
    \leavevmode
    \includegraphics[width=0.8\linewidth]{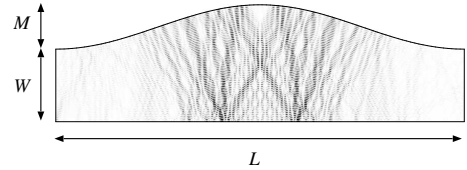}
    \caption{The cosine billiard for the parameters $W/L=0.18$ and
      $M/L=0.11$ with a gray-scale plot of the density of the resonant
      scattering state leading to the Husimi representation shown in Fig.
      \ref{fig:3} b).}
    \label{fig:1}
  \end{center}
\end{figure}

We study the cosine billiard as in Ref.~\cite{HucKetLew99} , either
closed (hard wall boundaries at $x=0$ and $x=L$) or with semi-infinite leads
attached to both sides. The boundary consists of the line $y=0$ and
\begin{equation}
  \label{eq:2}
  y(x)=W+\frac{M}{2}\left[1-\cos\left(\frac{2\pi x}{L}\right)\right]
\end{equation}
for $0\le x\le L$ (see Fig. 1). The classical phase space structure
can be tuned by varying the ratios $W/L$ and $M/L$. For $W/L=0.18$ and
$M/L=0.11$ the system has a mixed phase space. We take $E_0=\hbar^2
\pi^2 / (2 m W^2)$ as the unit of energy.

\begin{figure}[tbp]
  \begin{center}
    \leavevmode
    \includegraphics[width=1.0\linewidth]{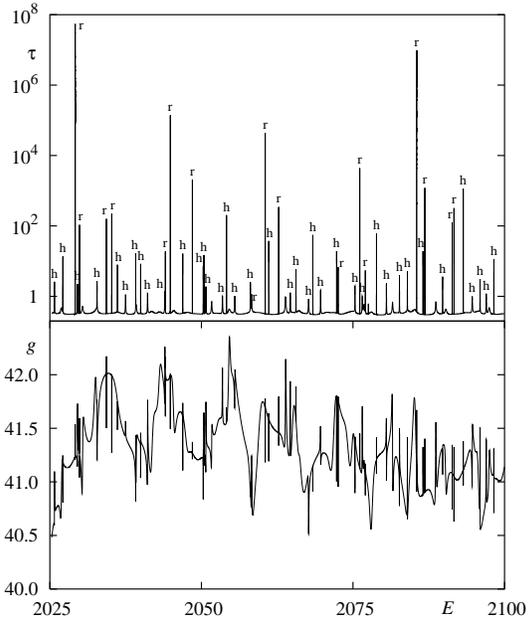}
    \caption{Wigner-Smith delay time $\tau$ (upper part) and
      conductance $G$ (lower part) vs energy $E$. In the upper part,
      the labels of the resonances indicate whether they correspond to
      regular (r) or hierarchical (h) eigenstates of the closed
      system.}
    \label{fig:2}
  \end{center}
\end{figure}
Fig. \ref{fig:2} shows the isolated resonances occuring in both the
conductance and the Wigner-Smith time delay $\tau=\frac{-i\hbar}{2N}
\mbox{Tr}\,(S^\dagger dS/dE)$ ($2N$ is the dimension of the
$S$-matrix). The calculations of $S$ and $\tau$ are outlined in
Refs.~\cite{HucKetLew99,bhk01}.

\begin{figure}[tbp]
  \begin{center}
    \leavevmode
    \includegraphics[width=0.8\linewidth]{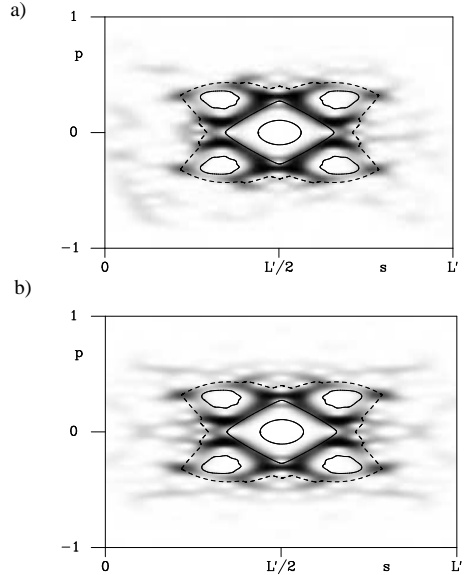}
    \caption{(a) Husimi representation of a hierarchical eigenstate of
      the closed system. (b) A similar phase space representation of
      the corresponding scattering state (see also Fig. \ref{fig:1}).
      Also shown are the most important Kolmogorov-Arnold-Moser tori
      (solid lines) and the main border to the chaotic region (dashed
      line).}
    \label{fig:3}
  \end{center}
\end{figure}

In order to elucidate the origin of the resonances, we have calculated
the associated scattering states of the open system as well as the
eigenstates of the corresponding closed system. Fig.~\ref{fig:3} shows
a comparison of the Husimi representation of the eigenstates and
scattering states with the classical phase space structure
\cite{bhk01}. The energy of the scattering state coincides with the
resonance energy 2041.109 and the eigenenergy of the eigenstate
differs from this energy by less than 10\% of the mean level spacing.
It is apparent that this resonance is due to an eigenstate with a
phase space representation concentrated on the hierarchical part of
phase space. Repeating this analysis for the other resonances allows
for them to be labeled as hierarchical or regular depending on the
part of phase space that their Husimi representations are concentrated on.
The resulting labels are shown in Fig.~\ref{fig:2}.

To conclude, we have analyzed phase space portraits of both
eigenstates of a closed billiard and scattering states of the opened
billiard. This allowed us to identify the origin of sharp, isolated
resonances in the conductance and Wigner-Smith time of the billiard
and to classify the resonant states as either regular or hierarchical. In
addition, we saw that the definition of hierarchical states of Ref.
\cite{KetHufSteWei2000} can be transferred to scattering states as
well.
%
%
\vspace{-0.5cm}
\begin{ack}
A.B.\ acknowledges support by the 
Deutsche Forschungs\-ge\-mein\-schaft under contract No. DFG-Ba 1973/1-1.
\end{ack}
\vspace{-0.2cm}
%
%


\begin{thebibliography}{9}
\bibitem{Jal2000}
R.~A. Jalabert,  in: Proceedings
  of the International School of Physics ``Enrico Fermi'', Course CXLIII {\it
  New Directions in Quantum Chaos}, G.~Casati, I.~Guarneri and U.~Smilansky
  (eds.), IOS Press Amsterdam (2000).

\bibitem{LicLie92}
A.~J. Lichtenberg, M.~A. Liebermann, {\em Regular and chaotic dynamics,},
  2nd  ed. (Springer-Verlag, New York, 1992).

\bibitem{Ket96}
R. Ketzmerick, Phys. Rev. B {\bf 54}  (1996)  10841.

\bibitem{HucKetLew99}
B. Huckestein, R. Ketzmerick, C.~H. Lewenkopf, Phys. Rev. Lett. {\bf 84}
  (2000)  5504; \emph{ibid.} {\bf 87} (2001) 119901(E); and
  references therein.

\bibitem{KetHufSteWei2000}
R. Ketzmerick, L. Hufnagel, F. Steinbach, M. Weiss, Phys. Rev. Lett. {\bf
  85}  (2000)  1214.

\bibitem{HufWeiKet2001}
L. Hufnagel, M. Weiss, R. Ketzmerick, Europhys. Lett. {\bf 54}
  (2001)  703.

\bibitem{bhk01}
A. B\"acker, A. Manze, B. Huckestein, R. Ketzmerick, Phys. Rev. E art. no. 016211 (2002).

\end{thebibliography}
\end{document}